\documentclass[lettersize,journal]{IEEEtran}
\usepackage{amsmath,amsfonts}
\usepackage{algorithmic}
\usepackage{algorithm}
\usepackage{array}
\usepackage[caption=false,font=normalsize,labelfont=sf,textfont=sf]{subfig}
\usepackage{textcomp}
\usepackage{stfloats}
\usepackage{url}
\usepackage{hyperref}
\usepackage{verbatim}
\usepackage{graphicx}
\usepackage{cite}
\usepackage{tikz}
\usepackage{pgfplots}
\usepackage[dvipsnames]{xcolor}

\pgfplotsset{width=8.4cm,compat=1.18}
\pgfplotsset{every x tick label/.append style={font=\small, yshift=0.5ex},every y tick label/.append style={font=\small, xshift=0.5ex},
every axis legend/.append style={
at={(5,1)},
anchor=north west,font=\large
}}

\definecolor{clr1}{rgb}{0.0, 0.0, 1.0}
\definecolor{clr2}{rgb}{0.96, 0.73, 1.0}
\definecolor{clr3}{rgb}{1.0, 0.01, 0.24}
\definecolor{clr4}{rgb}{0.0, 0.5, 0.0}
\definecolor{clr5}{rgb}{1.0, 0.49, 0.0}
\definecolor{clr6}{rgb}{0.1, 0.1, 0.44}
\begin{document}

\title{Sequential Processing Strategies in Fronthaul Constrained Cell-Free Massive MIMO Networks}

\author{Vida Ranjbar,~\IEEEmembership{Member,~IEEE,} Robbert Beerten,~\IEEEmembership{Graduate Student Member,~IEEE,} Marc Moonen,~\IEEEmembership{Fellow,~IEEE,} and Sofie Pollin,~\IEEEmembership{Senior Member,~IEEE}
% \IEEEauthorblockA{
%  \\
% Email: \{robbert.beerten, marc.moonen, sofie.pollin\}@kuleuven.be, vida.ranjbar@aau.at
% }
        % <-this % stops a space
\thanks{Corresponding author: Vida Ranjbar (vida.ranjbar8@gmail.com) is with the Institute of Networked and Embedded Systems, University of Klagenfurt, Austria. Robbert Beerten, Marc Moonen and  Sofie Pollin are with the department of Electrical Engineering (ESAT), KU Leuven, Belgium (Email: robbert.beerten, marc.moonen, sofie.pollin\}@kuleuven.be). This work is funded by the European  Union’s Horizon 2020 research
and innovation program under grant agreement 101096954 (6G-BRICKS)
, and Research Foundation –
Flanders (FWO) under project number G0C0623N. The resources and services used in this work were provided by the VSC
(Flemish Supercomputer Center), funded by the Research Foundation—
Flanders (FWO) and the Flemish Government.
% Link to the MATLAB code for this paper will be published upon publication.
}% <-this % stops a space
%\thanks{Manuscript received April 19, 2021; revised August 16, 2021.}
}

% \markboth{IEEE Wireless Communications Letters}%


\maketitle

\begin{abstract}
In a cell-free massive MIMO (CFmMIMO) network with a daisy-chain fronthaul, the amount of information that each access point (AP) needs to communicate with the next AP in the chain is determined by the location of the AP in the sequential fronthaul. Therefore, we propose two sequential processing strategies to combat the adverse effect of fronthaul compression on the sum of users' spectral efficiency (SE): 1) linearly increasing fronthaul capacity allocation among APs and 2) Two-Path users' signal estimation. The two strategies show superior performance in terms of sum SE compared to the equal fronthaul capacity allocation and Single-Path sequential signal estimation.
\end{abstract}
\section{Introduction}
Sequential processing in CFmMIMO networks has attracted significant research attention in recent years, as evidenced by numerous studies such as
\cite{shaik2020,shaik2021,Ioanis_ISNCC_2023,Ioaniss_TWC_2024,Jo_joint_prc_comp} among others. In a sequential network, the APs take turns in refining users' signal estimate in the uplink direction. The architecture reduces the total fronthaul signaling among the APs and the CPU and has a low practical implementation complexity \cite{shaik2021}, which makes it the favored architecture in some particular scenarios such as railways, factories, museums, etc. Using Kalman filtering, it can be proved that the performance of a distributed sequential uplink signal estimation at the APs is the same as the performance of the centralized linear minimum mean square error (LMMSE) signal estimation \cite{shaik2021,ke_Dmimo_kalman}, under ideal condition such as no fronthaul compression requirement.

 Fronthaul signaling among APs/CPU over limited-capacity (wired or wireless) links constitutes an integral part of the whole CFmMIMO networking. 
The performance degradation due to  the limited-capacity fronthaul links
%The limitation that brings to network performance 
can be mitigated in several ways \cite{Ioanis_meditcom_2022,Ioanis_ISNCC_2023,Ioaniss_TWC_2024,Ioanis_spawc_2024}, among others.
In a sequential architecture, the fact that APs can exchange information among each other can be exploited 
to indirectly lower the bit length of the exchanged users' signal estimates \cite{VRAN_eucnc_2024}.
Different strategies, such as compare and forward (CNF) \cite{Ioanis_ISNCC_2023, Ioaniss_TWC_2024}, sequential mean squared error (MSE) minimization-based user-AP association \cite{Ioanis_spawc_2024} 
and sequential maximization of the local mutual information between the users’ signals and information available at each AP under limited-capacity fronthaul links \cite{Ioannis_2024_MI}, are considered to mitigate the adverse effect of a limited-capacity fronthaul link on users' SE.
The authors in \cite{zhang2020downlink}, considered the downlink CFmMIMO with serial fronthaul and took an information-theoretic approach towards fronthaul capacity allocation among APs. They demonstrated that having a total fronthaul capacity, water-pouring distribution of the total fronthaul capacity over fronthaul links among APs will optimize the channel capacity.

\textbf{Motivation:} 
In a sequential CF-mMIMO architecture, the amount of information an AP sends to the next AP depends on its position in the fronthaul chain. AP $l$ refines the users' signal estimates received from AP $l-1$ (i.e. $\tilde{\mathbf{s}}_{l-1}$) using its local received vector $\mathbf{y}_{l}$, and then forwards the compressed refined users’ signal estimate $\tilde{\mathbf{s}}_{l}$ to AP $l+1$. The estimate thus passes through APs $1,\dots,l$ before it reaches AP $l+1$. Consequently, each AP must convey the information embedded in all preceding APs’ received signals to the next AP/CPU. As the index $l$ increases, the forwarded estimate $\tilde{\mathbf{s}}_{l}$ reflects processing over an increasing number of antennas. Therefore, allocating equal fronthaul capacity to all APs is inefficient, since early APs may not use their full capacity while later APs require more to forward their refined estimates with lower compression noise.

\textbf{Contributions}: In this paper, we study the uplink sum SE performance of a practical limited-capacity sequential fronthaul CFmMIMO network.
%and challenge the equal fronthaul capacity allocation to all APs in the sequence, despite their index in the sequence.
We propose two fronthaul deployment strategies that can significantly improve the sum SE. %Subsequently, we propose 
The first strategy considers non-uniform linearly increasing fronthaul allocation among APs under the assumption of a fixed total number of bits per uplink sample for fronthaul signaling. One application scenario for linearly increasing fronthaul capacity is when the fronthaul links between APs are wireless, and there is a particular frequency bandwidth that is allocated for wireless fronthaul signaling. 
% To avoid inter-AP interference in fronthaul signaling, the frequency reuse factor is one.
The amount of spectrum that is allocated to each AP for fronthaul signaling increases with the AP index. Finally, a combination of Two-path signal processing with linearly increasing fronthaul gives the highest sum SE, especially for a higher number of users. Matlab source code can be found in \url{https://gitlab.com/VidaRanjbar/Sequential-fronthaul-constrained-CFmMIMO_WCL2026}.

\textbf{Notations}: We denote vectors and matrices with boldface lower-case and upper-case letters, respectively. Transpose and conjugate transpose operations are denoted by $^{\text{T}}$ and $^{\text{H}}$, respectively. A circularly symmetric complex Gaussian distribution with covariance matrix $\mathbf{X}$ is represented as $\mathcal{C}\mathcal{N}(0, \mathbf{X})$. Symbol $\mathbb{E}\{\mathbf{x}\}$ denotes the mean of $\mathbf{x}$. $diag(\mathbf{x})$ is a diagonal matrix with the same diagonal elements as the elements of vector $\mathbf{x}$. $\mathbf{X}\succeq\mathbf{0}$ means $\mathbf{X}$ is positive semi-definite.
\section{System model}\label{paper5_sysmodel}
This paper considers the uplink of a CFmMIMO network with APs connected in a sequential fronthaul, as shown in Fig. \ref{bidir_sig_procc}. There are $K$ single-antenna users in the network served by $L$ APs, each equipped with $N$ antennas. The total number of antennas is $M=LN$. AP $l$ refines the compressed users' signal estimates received from AP $l-1$, i.e., $\tilde{\mathbf{s}}_{l-1}$, based on its own received signal vector in the uplink, formulated as:
    \begin{equation}   \mathbf{y}_l=\mathbf{H}_l\mathbf{s}+\mathbf{n}_l,
        \label{rec_sig_APl}
    \end{equation}

where $\mathbf{s}\sim \mathcal{CN}(\mathbf{0},p\mathbf{I}_K)$ is users' signal vector, $\mathbf{n}_l\sim\mathcal{CN}(\mathbf{0},\sigma^2\mathbf{I}_N)$ is the noise vector at AP $l$. Each column of matrix $\mathbf{H}_l$ follows a circularly symmetric complex Gaussian distribution, i.e., $\mathbf{H}_{l[:,k]}\sim\mathcal{CN}(\mathbf{0},\mathbf{R}_{kl})$.
We don't elaborate on channel estimation and refer to \cite{massivemimobook} for the channel estimation procedure. It is assumed that the local channel realizations are known at the APs. The local channel realizations remain constant during one coherence block of time and frequency, with a total $\tau_c$ sample per block. The phase and carrier frequency synchronization between distributed APs to enable coherent transmission and reception in time-division duplex (TDD) CFmMIMO is discussed in \cite{beamsync}. In this paper, perfect synchronization is assumed among APs.%for the following signal estimation procedure.

The two mentioned sources of information, i.e., the users' signal estimate received from AP $l-1$ and the local received signal vector at AP $l$, are combined based on the local channel matrix, i.e., $\mathbf{H}_l$, and the error statistics of the estimates received from AP $l-1$, i.e., $\mathbf{C}_{l-1}=\mathbb{E}\{\mathbf{e}_{l-1}\mathbf{e}_{l-1}^{\text{H}}\}$, $\mathbf{e}_{l-1}=\mathbf{s}-\tilde{\mathbf{s}}_{l-1}$.
$\tilde{\mathbf{s}}_{l-1}$ is formulated as follows: 
\begin{equation}
    \tilde{\mathbf{s}}_{l-1}=\hat{\mathbf{s}}_{l-1}+\mathbf{q}_{l-1},
    \label{WCL2025_compr_refined_signal}
    \end{equation}
where $\hat{\mathbf{s}}_{l-1}$ is the refined users' signal estimate at AP $l-1$ before compression, and $\mathbf{q}_{l-1}$ is the compression noise vector at AP $l-1$, independent from $\hat{\mathbf{s}}_{l-1}$ and imposed by the limited-capacity fronthaul link that connects AP $l-1$ to AP $l$. Based on standard results of information theory in \cite[Ch. 3]{elgamal_kim_2011}, $\mathbf{q}_l\sim \mathcal{CN}(\mathbf{0}, \mathbf{Q}_l)$. Using Kalman filtering concept \cite{shaik2021,ke_Dmimo_kalman}, the refined signal estimate at AP $l$ is as follows:
\begin{equation}
\begin{aligned}
    \hat{\mathbf{s}}_{l}&=\tilde{\mathbf{s}}_{l-1}+\mathbf{\Gamma}_l(\mathbf{y}_l-\mathbf{H}_l\tilde{\mathbf{s}}_{l-1})\\&=\tilde{\mathbf{s}}_{l-1}+\mathbf{\Gamma}_l(\mathbf{H}_l\mathbf{e}_{l-1}+\mathbf{n}_l),
\end{aligned}
\label{WCL2025_refined_sig_est}
\end{equation}
where $\mathbf{\Gamma}_l$ is the $K\times N$ combining matrix computed as follows:
\begin{equation}
 \mathbf{\Gamma}_l=\mathbf{C}_{l-1}\mathbf{H}_l^{\text{H}}(\mathbf{H}_l\mathbf{C}_{l-1}\mathbf{H}^{\text{H}}_l+\sigma^2\mathbf{I}_N)^{-1}. 
 \label{WCL2025_Gamma}
\end{equation} 
At AP $1$, $\mathbf{C}_0=p\mathbf{I}_K$ and $\tilde{\mathbf{s}}_0=\mathbf{0}_{K\times 1}$.
 Based on (\ref{WCL2025_compr_refined_signal}) and (\ref{WCL2025_refined_sig_est}), $\tilde{\mathbf{s}}_l$ can be formulated as follows:
\begin{equation}
\begin{aligned}
 \tilde{\mathbf{s}}_{l}&
 %=\tilde{\mathbf{s}}_{l-1}+\mathbf{\Gamma}_l(\mathbf{y}_l-\mathbf{H}_l\tilde{\mathbf{s}}_{l-1})\\&
 =\hat{\mathbf{s}}_l+\mathbf{q}_l\\&=\begin{bmatrix}
     \mathbf{I}_K-\mathbf{\Gamma}_l\mathbf{H}_l&&\mathbf{\Gamma}_l
 \end{bmatrix}\begin{bmatrix}
     \hat{\mathbf{s}}_{l-1}\\\mathbf{y}_l
 \end{bmatrix}+(\mathbf{I}_K-\mathbf{\Gamma}_l\mathbf{H}_l)\mathbf{q}_{l-1}+\mathbf{q}_l.
\label{WCL2025_Compr_refined_sig_est1}  
\end{aligned}
\end{equation}
By recursion, $\tilde{\mathbf{s}}_{l}$ at AP $l$ can be rewritten as the combination of received signal vectors of AP $1$ to AP $l$ and an additive collective compression noise vector that was added to the users' signal estimates through the sequence until AP $l$:
%\robbert{I would consider removing this first subequation and adding the underbrace to the second part. You can save a lot of space here
%}
\begin{equation}
\begin{aligned}
\tilde{\mathbf{s}}_{l}&=\begin{bmatrix}\mathbf{V}_{1l}&&\cdots&&\mathbf{V}_{ll}
\end{bmatrix}\begin{bmatrix}\mathbf{y}^T_1&&\cdots&&\mathbf{y}^T_{l}
\end{bmatrix}^T\\&+\underbrace{\begin{bmatrix}\mathbf{A}_{1l}&&\cdots&&\mathbf{A}_{ll}
\end{bmatrix}\begin{bmatrix}\mathbf{q}^T_1&&\cdots&&\mathbf{q}^T_{l}
 \end{bmatrix}^T}_{\mathbf{q}_{l}^{\text{eff}}}
%\label{refined_sig_est_2}
\\&=\sum_{i=1}^l  \mathbf{V}_{il}\mathbf{H}_i\mathbf{s}+\mathbf{V}_{il}\mathbf{n}_i+\mathbf{A}_{il}\mathbf{q}_i,
\label{refined_sig_est_2}
\end{aligned}
\end{equation}
where 
\begin{equation}
\mathbf{V}_{il}=
\begin{cases}
\left(\prod_{j=i+1}^l(\mathbf{I}_K-\mathbf{\Gamma}_j\mathbf{H}_j)\right)\mathbf{\Gamma}_i, & \text{for } i < l\\
    \mathbf{\Gamma}_l, & \text{for } i = l \\
  \end{cases}
  \label{sigcombiner}
\end{equation} 
and
\begin{equation}
\mathbf{{A}}_{il}=
\begin{cases}
    \prod_{j=i+1}^l(\mathbf{I}_K-\mathbf{\Gamma}_j\mathbf{H}_j), & \text{for } i < l\\
    \mathbf{I}_K, & \text{for } i = l \\
  \end{cases}
  \label{Ncombiner}
\end{equation}
%Note that the matrices $\mathbf{A}_i \forall i$ and $\mathbf{V}_i, \forall i$ are dependent to the AP index. However, the dependency is to shown with an additional  
where the order of matrix multiplications is AP $l$, AP $l-1$, $\hdots$, AP $i+1$ from left to right. It is easily shown that $\mathbf{V}_{il}=\{(\mathbf{I}_K-\mathbf{\Gamma}_l\mathbf{H}_l)\mathbf{V}_{i(l-1)}, \forall i\leq l-1\}$ and $\mathbf{A}_{il}=\{(\mathbf{I}_K-\mathbf{\Gamma}_l\mathbf{H}_l)\mathbf{A}_{i(l-1)}, \forall i\leq l-1\}$.
Finally, the error covariance matrix on $\tilde{\mathbf{s}}_{l}$ is formulated as follows:
\begin{equation}
\begin{aligned}
    \mathbf{C}_l&=\mathbb{E}\{\mathbf{e}_l\mathbf{e}_l^{\text{H}}\}\\&=\mathbb{E}\{(\mathbf{s}-\tilde{\mathbf{s}}_l)(\mathbf{s}-\tilde{\mathbf{s}}_l)^{\text{H}}\}\\&=(\mathbf{I}_K-\mathbf{\Gamma}_l\mathbf{H}_l) \mathbf{C}_{l-1}+\mathbf{Q}_l,
    \label{WCL_est_errr_post_compr}
    \end{aligned}
\end{equation}
which can be proved considering (\ref{WCL2025_refined_sig_est}), (\ref{WCL2025_Gamma}) and (\ref{WCL2025_Compr_refined_sig_est1}).
%\subsection{Spectral efficiency }
Along with the uplink samples, the covariance matrices $\mathbf{C}_{l-1}$ should also be exchanged with AP $l$. However, as this matrix will remain constant over one coherence block and changes infrequently, we assume it is exchanged with high precision.
If the sequential processing of the signal of User $k$ is terminated at AP $l$ and the rest of the APs only relay the estimate toward the CPU, an achievable SE for user $k$ can be computed as follows:
\begin{equation}
    \text{SE}_{kl}=\frac{\tau_u}{\tau_c}\mathbb{E}\{\log_2(1+\text{SINR}_{kl})\}.
    \label{se}
\end{equation}
where $\text{SINR}_{kl}$ is defined in (\ref{SINR}). $\tau_c$ and $\tau_u$ are the number of samples per coherence block and number of samples for uplink data transmission per coherence block, respectively.
\begin{figure*}[t]
\begin{equation}
     \begin{aligned}
\text{SINR}_{kl}&=&\frac{p|\sum\limits_{i=1}^l\mathbf{V}_{il[k,:]}\mathbf{H}_{i[:,k]}|^2}{\underbrace{\sum\limits_{j\neq k}p|\sum\limits_{i=1}^l\mathbf{V}_{il[k,:]}\mathbf{H}_{i[:,j]}|^2+\sigma^2\sum\limits_{i=1}^l\mathbf{V}_{il[k,:]}\mathbf{V}_{il[k,:]}^{\text{H}}+\sum\limits_{i=1}^{l-1}\mathbf{A}_{il[k,:]}\mathbf{Q}_i\mathbf{A}_{il[k,:]}^{\text{H}}+\mathbf{Q}_{l[k,k]}}_{X_{kl}}}.
     \end{aligned}
    \label{SINR}
 \end{equation}
 \end{figure*}
\section{Fronthaul compression noise}\label{sec_fc}
As already mentioned, for a fixed fronthaul capacity at a particular AP, the user's signal estimates are refined, compressed, and then sent to the next AP. In this paper, we use equal inter-user (EIU) fronthaul compression where a particular AP allocates the same number of bits to each user's signal. This is an element-wise compression scheme. 
%\robbert{What exactly do you mean by equal bit compression? Does it resemble the number of bits for each fronthaul? }
Alternatively, we consider fronthaul compression based on sum compression noise variance minimization (SCNM), formulated as minimization of the trace of the compression noise covariance matrix in \cite{Ioaniss_TWC_2024}. Lastly, we introduce a weighted sum interference plus noise variance minimization-based fronthaul compression method (WSINM). The last two compression methods are categorized as vector-wise compression.

Based on (\ref{WCL2025_compr_refined_signal}), (\ref{WCL2025_refined_sig_est}), (\ref{WCL2025_Gamma}), and (\ref{WCL_est_errr_post_compr}), the correlation matrix of $\hat{\mathbf{s}}_l$ (refined signal before compression at AP $l$) is defined as:
\begin{equation}
\begin{aligned}
   \mathbf{P}_l&=\mathbb{E}\{\hat{\mathbf{s}}_l\hat{\mathbf{s}}^{\text{H}}_l|\mathbf{C}_{l-1}, \mathbf{H}_l\}\\&=
   \mathbf{P}_{l-1}+\mathbf{Q}_{l-1}+\mathbf{\Gamma}_l\mathbf{H}_l\mathbf{C}_{l-1}-\mathbf{Q}_{l-1}\mathbf{H}_l^\text{H}\mathbf{\Gamma}^\text{H}_l-\mathbf{\Gamma}_l\mathbf{H}_l\mathbf{Q}_{l-1}.
   \end{aligned}
\end{equation}
The derivations are not detailed due to space limitations. However, it is easily verifiable. Note that $\mathbb{E}\{\tilde{\mathbf{s}}_{l-1}\mathbf{e}_{l-1}^\text{H}\}=-\mathbf{Q}_{l-1}$. 
    \subsection{Element-wise compression: Equal inter-user fronthaul capacity allocation}
 If there are $b_{kl}$ bits to compress the $k^{th}$ complex valued element of $\hat{\mathbf{s}}_{l}$  at AP $l$, based on standard results of rate distortion theory \cite[Ch.3]{elgamal_kim_2011}, the relation between $b_{kl}$ and the $k^{\text{th}}$ diagonal element of matrices $\mathbf{P}_l$ and $\mathbf{Q}_l$ is as follows:
\begin{equation}
    b_{kl}=\log_2(\mathbf{P}_{l[k,k]}\mathbf{Q}_{l[k,k]}^{-1}+1)\rightarrow \mathbf{Q}_{l[k,k]}=\frac{\mathbf{P}_{l[k,k]}}{2^{b_{kl}}-1}.
\end{equation}
In the simulation section, we consider $b_{kl}=R_l/K$ for EIU, with $R_l$ being the number of bits/sample allocated to AP $l$ for fronthaul signaling of $\hat{\mathbf{s}}_l$.
\subsection{Vector-wise compression: Sum compression noise variance minimization}
To compute the compression noise covariance matrix $\mathbf{Q}_l$ at a particular AP $l$, a sum compression noise variance minimization (SCNM) problem in \cite{Ioaniss_TWC_2024} is formulated as follows:
\begin{equation}
     \begin{aligned}
\arg \min_{\mathbf{Q}_{l}\succeq\mathbf{0}} \quad  & \sum_{k} \mathbf{Q}_{l[k,k]}\\
\textrm{s.t.} \quad & R_l=\log_2\det(\mathbf{P}_l\mathbf{Q}_l^{-1}+\mathbf{I}_K).
\end{aligned} 
\label{min_quantpower}
\end{equation}
Problem (\ref{min_quantpower}) can be solved using Proposition 1 in \cite{Ioaniss_TWC_2024}.

%\robbert{Perhaps it would be useful to discuss the trade-off in required compute power. I believe that for solving this problem, you need $LK$ bisection searches for the dual variables, right? }
%The solution to the optimization problem in \ref{min_quantpower} is elaborated in \cite{}
\subsection{Vector-wise compression: Weighted sum interference plus noise variance minimization}
Alternatively, inspired by the weighted MMSE (WMMSE) problem in \cite{wmse_2011}, a weighted sum interference plus noise variance minimization (WSINM) problem is formulated to compute the compression noise covariance matrix, as follows:

\begin{equation}
     \begin{aligned}
\arg \min_{\mathbf{Q}_{l}\succeq\mathbf{0},w_{lk} \forall k} \quad  & \sum_{k} w_{lk}X_{lk}-\sum_{k} \log_2 w_{lk}\\
\textrm{s.t.} \quad & R_l=\log_2\det(\mathbf{P}_l\mathbf{Q}_l^{-1}+\mathbf{I}_K).
\end{aligned} 
\label{min_intf_prob_VC}
\end{equation}
where $X_{kl}$ is defined in (\ref{SINR}). The above problem is not jointly convex with respect to $\mathbf{Q}_l$ and $\{w_{lk}, \forall k\}$. Hence, the block coordinate descent (BCD) procedure is used to iteratively optimize the multi-variable function, as follows:
\begin{itemize}
    \item For a fixed $\{w_{lk}, \forall k\}$, the problem (\ref{min_intf_prob_VC}) reduces to the minimization of a weighted trace of the compression noise covariance matrix that can be solved similarly to the minimization strategy in problem (\ref{min_quantpower}). More specifically, for a fixed $\{{w}_{lk}, \forall k\}$, matrix $\mathbf{W}_l$ is defined as:
\begin{equation}
    \mathbf{W}_l=diag(w_{l1},\cdots,w_{lK}),
\end{equation}
and the problem in (\ref{min_intf_prob_VC}) is equivalent to:
\begin{equation}
     \begin{aligned}
\arg & \min_{\mathbf{Q}_{l}\succeq\mathbf{0}}    \sum_{k=1}^K w_{lk}\mathbf{Q}_{l[k,k]}\quad=tr(\mathbf{W}_l\mathbf{Q}_{l})\\ &=tr(\underbrace{\mathbf{W}_l^{\frac{1}{2}}\mathbf{Q}_{l}\mathbf{W}_l^{\frac{\text{H}}{2}}}_{\bar{\mathbf{Q}}_l})\\
\textrm{s.t.} \quad & R_l=\log_2\det(\underbrace{\mathbf{W}_l^{\frac{1}{2}}\mathbf{P}_l\mathbf{W}_l^{\frac{\text{H}}{2}}}_{\bar{\mathbf{P}}_l}\underbrace{\mathbf{W}_l^{\frac{-1}{2}}\mathbf{Q}_l^{-1}\mathbf{W}_l^{\frac{-\text{H}}{2}}}_{\bar{\mathbf{Q}}_l^{-1}}+\mathbf{I}_K),
\end{aligned} 
\label{min_intf_prob_VC_fixed_w}
\end{equation}
where the properties of the trace, determinant, and the square root of a matrix are used \cite{mcook}.
\item For a fixed $\mathbf{Q}_l$ the problem in (\ref{min_intf_prob_VC}) is concave with respect to $\{w_{lk}, \forall k\}$ and the optimal $\{w_{lk}, \forall k\}$ are calculated as:
\begin{equation}
    w_{lk}^*=\frac{1}{\ln2\times X_{lk} }, \forall k.
\end{equation}
\end{itemize}
The iteration continues until convergence. The convergence proof of the iterative optimization is straightforward using the fact that the objective function is decreasing in each iteration and it is bounded from below for $\{X_{lk}\neq0,\forall k\}$, which is almost always the case in practice. In other words, in practice, the interference plus noise is always non-zero for all users in a multi-user CFmMIMO network. See the Appendix for details.

If $\{w_{lk}^*, \forall k\}$ are inserted back into the problem (\ref{min_intf_prob_VC}), we have an equivalent problem as follows:
\begin{equation}
     \begin{aligned}
\arg \max_{\mathbf{Q}_{l}\succeq\mathbf{0}} \quad  & \sum_{k}\log_2 \frac{1}{X_{kl} }\\
\textrm{s.t.} \quad & R_l=\log_2\det(\mathbf{P}_l\mathbf{Q}_l^{-1}+\mathbf{I}_K).
\end{aligned} 
\label{min_intf_prob_eq}
\end{equation}
Based on (\ref{se}) and (\ref{SINR}), the objective function in (\ref{min_intf_prob_eq}) differs from sum SE up to a constant (the numerator in (\ref{SINR})) in the high SINR regime. Hence, (\ref{min_intf_prob_eq}) is equivalent to the sum SE maximization problem in that regime.

\section{Fronthaul capacity allocation}\label{sec_IV}
In an uplink sequential signal estimation without any constraint on the capacity of the fronthaul links and the need for compression, the number of bits forwarded from one AP to the next increases with AP index.
 In the presence of limited-capacity fronthaul, the APs in the beginning need less capacity than the APs toward the end of the sequence. If there is a total fronthaul capacity (for example, there is a fixed number of subcarriers allocated to wireless fronthaul signaling), then the capacity allocation to each AP should take into account the increasing need for fronthaul signaling capacity toward the end of the sequence. This means that equal capacity allocation to the APs might be suboptimal, especially when the number of users is large, as the first AP may not need the capacity allocated to it, while the last AP may need to compress the users' signals estimate heavily. We compare two fronthaul capacity allocations among APs:
\begin{itemize}
    \item Equal fronthaul allocation (EF): If the total number of bits per uplink sample of users' signal is $R_T$, the bit allocated to each AP is $R_l=\frac{R_T}{L}$.
    \item Linearly increasing fronthaul capacity allocation (LF): In this approach, the fronthaul capacity increases linearly with AP index, $R_{l}=\frac{2R_T}{L(L+1)}l$. 
\end{itemize}
The fronthaul capacity allocation scheme can also be logarithmic, i.e., $R_{l}=\frac{\log_2 l}{\sum_i\log_2 i}R_T$. This capacity allocation strategy typically exhibits inferior performance compared to the linearly increasing fronthaul capacity allocation in the considered scenario and is therefore removed from the simulation section.
\section{Two-Path sequential signal propagation}
Based on (\ref{WCL2025_Compr_refined_sig_est1}), the signal estimate in AP $l$ has the combined effect of compression noise from AP $1$ to AP $l$ as well as the corresponding received signal vectors. Hence, the longer the sequence, the bigger the combined effect of compression noise on the users' signal estimate as well as the amount of information to be exchanged with the next AP.

\begin{figure}
    \centering
\includegraphics[width=.7\linewidth]{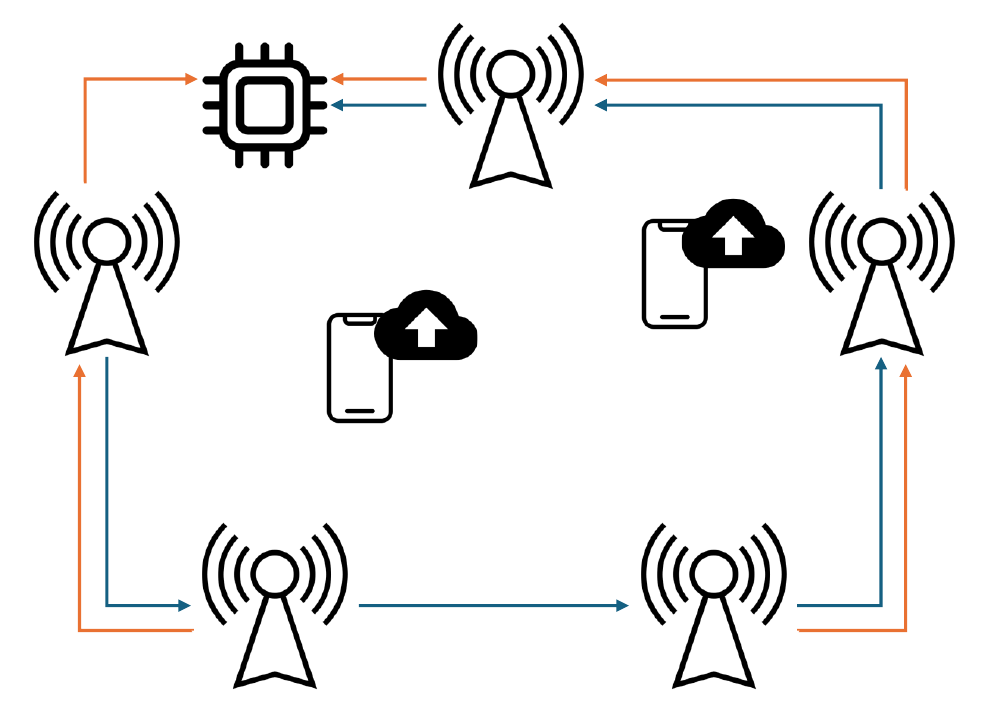}
    \caption{Sequential uplink signal estimation. User signal estimates are refined following \textbf{either} the blue path until it reaches CPU (Single-Path (SP) signal estimation), \textbf{or} it is refined following two short orange paths, and the two resulting signal estimates from the two paths get fused in the CPU (Two-Path (TP) signal estimation).
    }
    \label{bidir_sig_procc}
\end{figure}

%\robbert{Do you mean that the compression noise grows? Or that the relative effect of the compression noise is much bigger because of the increased quality of the signal estimate? }
To compensate for this combined effect of compression noise, the linear increase in the fronthaul capacity along the sequence is proposed in Section \ref{sec_IV}. An alternative to increasing the fronthaul capacity proposal is the Two-Path signal propagation, as shown in Fig. \ref{bidir_sig_procc}.
% \footnote{This figure has been designed using resources from Flaticon.com}.

The effect of increasing compression noise when fronthaul capacity is not growing sufficiently can be mitigated if the users' signal estimate propagates through two (instead of one) paths to the CPU, and then in the CPU, the local estimates from the two paths can be combined to estimate users' signals globally. In this case, the combined compression noise effect in each AP is at most from $L/2$ preceding APs instead of $L$ APs (with the assumption of equally long paths).
From Path $r\in\{1,2\}$, the CPU receives the following signal estimate:
    \begin{equation}
\begin{aligned}
\tilde{\mathbf{s}}^r&=\begin{bmatrix}\mathbf{V}^r_{1}&&\cdots&&\mathbf{V}^r_{L_r}
\end{bmatrix}\begin{bmatrix}\mathbf{y}^{rT}_1&&\cdots&&\mathbf{y}^{rT}_{L_r}
\end{bmatrix}^T\\&+\underbrace{\begin{bmatrix}\mathbf{A}^r_{1}&&\cdots&&\mathbf{A}^r_{L_r}
\end{bmatrix}\begin{bmatrix}\mathbf{q}^{rT}_1&&\cdots&&\mathbf{q}^{rT}_{L_r}
 \end{bmatrix}^T}_{\mathbf{q}^{\text{r-eff}}},
\label{refined_sig_est_3}  
\end{aligned}
\end{equation}
%\robbert{These indices are quite confusing to me since they are so similar to (7), and also for both fronthauls, they both start from index 1 up to $L_r$. Maybe you could introduce a notation like $L_r^l$ which could then resemble the $l$'th AP on the $r$'th fronthaul? }
where $L_r$ is the number of APs involved in sequential processing in Path $r$ and $L=\sum_{r=1}^2 L_r$.
%The signal reaching the CPU from Path $r$ is formulated as below:
% \begin{equation}
%     \tilde{\mathbf{s}}^r=\hat{\mathbf{s}}^r_{L_r}+\mathbf{q}^r_{L_r}
%     \label{compr_refined_signal_1}
%     \end{equation}
The effective channel and effective noise from path $r$ is defined as $\mathbf{G}^{r}=\sum_{l=1}^{L_r}\mathbf{V}^r_{l}\mathbf{H}^r_l$ and $\mathbf{z}^r=\sum_{l=1}^{L_r}\mathbf{V}^r_l\mathbf{n}^r_l+\mathbf{A}^r_l\mathbf{q}^r_l\sim\mathcal{CN}(\mathbf{0},\mathbf{Z}^r)$.
Covariance matrix $\mathbf{Z}^r$ is formulated as:
\begin{equation}
\mathbf{Z}^r=\sum_{l=1}^{L_r}\sigma^2\mathbf{V}^r_{l}\mathbf{V}^{r\text{H}}_{l}+\mathbf{A}^r_{l}\mathbf{Q}^r_l\mathbf{A}^{r\text{H}}_{l}.
\end{equation}
The superscript $r$ is meant to distinguish APs belonging to different paths. The overall information vector $\mathbf{y}^{\text{TP}}$ (TP stands for Two-Path), effective channel matrix $\mathbf{G}^{\text{TP}}$ and noise vector $\mathbf{z}^{\text{TP}}$ from two paths are formulated as follows respectively:
\begin{equation*}
    \mathbf{y}^{\text{TP}}=\begin{bmatrix}
\tilde{\mathbf{s}}^1\\\tilde{\mathbf{s}}^2
    \end{bmatrix},\quad \mathbf{G}^{\text{TP}}=\begin{bmatrix}
       \mathbf{G}^1\\\mathbf{G}^2 
    \end{bmatrix},\quad \mathbf{z}^{\text{TP}}=\begin{bmatrix}
       \mathbf{z}^1\\\mathbf{z}^2
    \end{bmatrix}\sim \mathcal{CN}(\mathbf{0},\mathbf{Z}^{\text{TP}}),
\end{equation*}
where $\mathbf{Z}^{\text{TP}}=blkdiag(\mathbf{Z}^1,\mathbf{Z}^2).$
The global LMMSE estimate of users' signal based on $\mathbf{y}^{\text{TP}}$ in the CPU is as follows:
\begin{equation}
\hat{\mathbf{s}}^{\text{TP}}=\mathbf{V}^{\text{TP}} \mathbf{y}^{\text{TP}},\quad \mathbf{V}^{\text{TP}} =p\mathbf{G}^{\text{TP}\text{H}}(p\mathbf{G}^{\text{TP}}\mathbf{G}^{\text{TP}\text{H}}+\mathbf{Z}^{\text{TP}})^{-1}.
\end{equation}
%where 
% \begin{equation}
%     \mathbf{V}^{\text{eff}} =p\mathbf{G}(p\mathbf{G}\mathbf{G}^{H}+\mathbf{N})^{-1}
% \end{equation}
\section{Simulation results}
We validate the proposed approaches for sequential signal estimation in a daisy chain CFmMIMO network. A total number of $M=120$ antennas are distributed among $L$ APs. The APs are located on the perimeter of a circle with a radius of $300m$ meters. The users are uniformly distributed in an inner circle with a radius of $150m$. The number of users and APs is specified under each simulation figure.
The large-scale fading coefficient between user $k$ and AP $l$ is as follows:
\begin{equation}
    \beta_{kl}=-30.5-36.7\log_{10}(d_{kl}),
\end{equation}
following the propagation model of 3GPP Urban Microcell \cite{3gpp_PL} with 2GHz carrier frequency and bandwidth of $B=100\text{MHz}$ \cite{3gpp_PL}. Parameter $d_{kl}$ is the distance between AP $l$ and user $ k$ in meters. The noise variance is assumed to be $\sigma^2=-85\text{dBm}$. The users' transmit power is set to $p=20\text{dBm}$ ($100\text{mWatts}$). It is assumed that $\tau_u=\tau_c-\tau_p$, where $\tau_c=200$ and $\tau_p=K$ samples are used in each coherence block to acquire channel knowledge. Suppose $R_T=1000 \text{bit/sample}$, it means that for one uplink transmission (one uplink sample compression), there are 1000 bits to be shared between APs. If there are $4$ APs, for example, in total, the linearly increasing fronthaul will allocate $R_1=100$, $R_2=200$, $R_3=300$ and $R_4=400$ bits to the APs $1$ to $4$. Afterwards, if the method in III-A is used for inter-user fronthaul capacity allocation, then AP $1$ will allocate $b_{k1}=100/K$ bits to each user, AP $2$ will allocate $b_{k2}=200/K$ bits to each user, etc. $R_l$ can be converted to the equivalent bandwidth of the link connecting AP $l$ to AP $l+1$=$R_l \times \#$ \text{uplink} \text{sample/second}.

\begin{figure}[ht]
    \centering
    \pgfplotsset{width=8cm,compat=1.18}
\pgfplotsset{every x tick label/.append style={font=\small},every y tick label/.append style={font=\small},
every axis legend/.append style={
at={(5,1)},
anchor=north west,font=\small
}}

%\definecolor{clr1}{rgb}{0.0, 0.0, 1.0}
%\definecolor{clr2}{rgb}{0.96, 0.73, 1.0}
%\definecolor{clr3}{rgb}{1.0, 0.01, 0.24}
%\definecolor{clr4}{rgb}{0.0, 0.5, 0.0}
%\definecolor{clr5}{rgb}{1.0, 0.49, 0.0}
%\definecolor{clr6}{rgb}{0.1, 0.1, 0.44}
%\pgfplotsset{width=7cm,compat=1.18}

%\begin{document}

%\begin{tikzpicture}
\begin{tikzpicture}[scale=0.87]

\begin{scope}[xshift=0cm,yshift=0cm]
\begin{axis}[
domain=0:4,
grid=major,
%ylabel= Average per-user SE (bit/sec/Hz),
xlabel= { Number of users},
xmin=1,
xmax=20,
xtick style={color=clr3},
 xticklabel style={yshift=-2.5pt},
ymin=0,
mark size=4.0 pt,
legend columns=3,
legend style={nodes={scale=0.9, transform shape},at={(.52,-0.17)},anchor=north},
ylabel={ Average sum SE (bit/sec/Hz)},
line width=1pt,
%yticklabels={5,6,7,8},
%legend entries={
%$Option 1$,$Option 2$,$Option 3$,$Option4$,$ OPtion 5$,
%{[text width=25pt,text depth=]Neg. Sign:},
%},
% same effect:
% legend style={
% nodes={text width=25pt,text depth=},}
]

        \addplot [red]
        table[x=R,y=WMSE_p1_eq,col sep=comma] {data/revision1/data_diffK_R500.csv};
       % \addlegendentry{EF-SP}
        \addplot [red, dashed]
        table[x=R,y=WMSE_p1_lin,col sep=comma] {data/revision1/data_diffK_R500.csv};
        %\addlegendentry{LF-SP}

        \addplot [OliveGreen]
        table[x=R,y=WMSE_p2_eq,col sep=comma] {data/revision1/data_diffK_R500.csv};
        %\addlegendentry{EF-TP}
        \addplot [OliveGreen, dashed]
        table[x=R,y=MSE_p2_lin,col sep=comma] {data/revision1/data_diffK_R500.csv};
       % \addlegendentry{LF-TP}

        \addplot [black]
        table[x=R,y=MMSE_cen_inff,col sep=comma] {data/revision1/data_diffK_R500.csv};
       % \addlegendentry{Infinite capacity}

% \end{axis}
% \end{scope}

% \begin{scope}[xshift=7cm,yshift=0cm]
% \begin{axis}[
% domain=0:4,
% %title={$K=64$},
% grid=major,
% %ylabel= Average per-user SE (bit/sec/Hz),
% xlabel= {\large Number of users},
% xmin=1,
% xmax=20,
% xtick style={color=clr3},
%  %xtick={1,2,3,4,5,6,7,8},
% % xticklabels={L=2,L=4,L=8,L=16,L=32,L=64,L=128,L=256},
%  xticklabel style={yshift=-2.5pt},
% ymin=0,
% %ymax=20,
% %ytick={4,5,6,7,8,9},
% mark size=4.0 pt,
% legend columns=5,
% legend style={nodes={scale=0.9, transform shape},at={(-.09,-0.25)},anchor=north},
% line width=1pt,
% yticklabels=\empty,
% %ylabel= Average sum spectral efficiency
% %yticklabels={5,6,7,8},
% %legend entries={
% %$Option 1$,$Option 2$,$Option 3$,$Option4$,$ OPtion 5$,
% %{[text width=25pt,text depth=]Neg. Sign:},
% %},
% % same effect:
% % legend style={
% % nodes={text width=25pt,text depth=},}
% ]

        \addplot [red]
        table[x=R,y=WMSE_p1_eq,col sep=comma] {data/revision1/data_diffK_R1000.csv};
        \addlegendentry{EF-SP}
        \addplot [red, dashed]
        table[x=R,y=WMSE_p1_lin,col sep=comma] {data/revision1/data_diffK_R1000.csv};
        \addlegendentry{LF-SP}

        \addplot [OliveGreen]
        table[x=R,y=WMSE_p2_eq,col sep=comma] {data/revision1/data_diffK_R1000.csv};
        \addlegendentry{EF-TP}
        \addplot [OliveGreen, dashed]
        table[x=R,y=WMSE_p2_lin,col sep=comma] {data/revision1/data_diffK_R1000.csv};
        \addlegendentry{LF-TP}

        \addplot [black]
        table[x=R,y=MMSE_cen_inff,col sep=comma] {data/revision1/data_diffK_R1000.csv};
        \addlegendentry{Infinite capacity}

        % \addplot [BurntOrange, mark=diamond]
        % table[x=R,y=MMSE_cen_inff,col sep=comma] {data/revision1/data_diffK_R1000.csv};
        % \addlegendentry{Infinite capacity}
        \draw[ black] (12,17) ellipse [x radius=0.5, y radius=6];
        
\draw[black] (16,31) ellipse [x radius=0.5, y radius=5];
    \draw[->] (12,17) -- (14,17) node[right] {$R_T=500$};    
    \draw[->] (16,31) -- (10,31) node[left] {$R_T=1000$};

\end{axis}
\end{scope}

\end{tikzpicture}

%\end{document}
    \caption{Impact of limited capacity fronthaul link: sum SE vs number of users. For bit allocation among users, WSINM is considered. $L=12$, $M=120$. 
    }
    \label{fig2_diffK}
\end{figure}
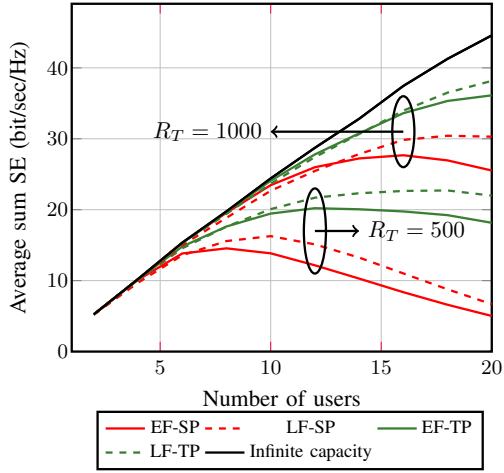

Fig. \ref{fig2_diffK} shows that when $K=20$ and $R_T=500$ bits/sample, linearly increasing fronthaul Two-Path signal estimation (LF-TP) and Equal fronthaul Two-Path signal estimation (EF-TP) will improve the sum SE performance by about $361\%$ and $262\%$ compared to the Equal Fronthaul Single-Path (EF-SP), respectively. 

\begin{figure}[ht]
    \centering
    \pgfplotsset{width=8cm,compat=1.18}
\pgfplotsset{every x tick label/.append style={font=\small},every y tick label/.append style={font=\small},
every axis legend/.append style={
at={(5,1)},
anchor=north west,font=\small
}}

%\definecolor{clr1}{rgb}{0.0, 0.0, 1.0}
%\definecolor{clr2}{rgb}{0.96, 0.73, 1.0}
%\definecolor{clr3}{rgb}{1.0, 0.01, 0.24}
%\definecolor{clr4}{rgb}{0.0, 0.5, 0.0}
%\definecolor{clr5}{rgb}{1.0, 0.49, 0.0}
%\definecolor{clr6}{rgb}{0.1, 0.1, 0.44}
%\pgfplotsset{width=7cm,compat=1.18}

%\begin{document}
%\begin{tikzpicture}
\begin{tikzpicture}[scale=0.87]
\begin{scope}
\begin{axis}[
domain=0:4,
grid=major,
xmin=100,
xmax=1000,
xtick style={color=clr3},
xticklabel style={yshift=-2.5pt},
xlabel= {$R_T$: Total fronthaul capacity (bit/uplink sample)},
ymin=0,
ymax=40,
mark size=4.0 pt,
legend columns=3,
legend style={nodes={scale=0.9, transform shape},at={(0.52,-0.2)},anchor=north},
ylabel= {Average sum SE (bit/sec/Hz)},
line width=1pt,
]

        \addplot [blue]
        table[x=R,y=eq_p1_eq,col sep=comma] {data/revision1/data_diffR.csv};
       % \addlegendentry{EF-SP}

        \addplot [BurntOrange]
        table[x=R,y=MSE_p1_eq,col sep=comma] {data/revision1/data_diffR.csv};
         %\addlegendentry{LF-SP}

        \addplot [red]
        table[x=R,y=WMSE_p1_eq,col sep=comma] {data/revision1/data_diffR.csv};
         %\addlegendentry{EF-TP}

         \addplot [blue, dashed]
        table[x=R,y=eq_p1_lin,col sep=comma] {data/revision1/data_diffR.csv};
       % \addlegendentry{EF-SP}

        \addplot [BurntOrange, dashed]
        table[x=R,y=MSE_p1_lin,col sep=comma] {data/revision1/data_diffR.csv};
        % \addlegendentry{LF-SP}

        \addplot [red, dashed]
        table[x=R,y=WMSE_p1_lin,col sep=comma] {data/revision1/data_diffR.csv};
         %\addlegendentry{EF-TP}
        %  \addplot [black,dashed]
        % table[x=R,y=eq_p1_log,col sep=comma] {data/revision1/data_diffR.csv};

        % \addplot [black, mark=diamond]
        % table[x=R,y=eq_p2_lin,col sep=comma] {data/revision1/data_diffR.csv};
        % \addlegendentry{LF-TP}

        %\addlegendentry{$b=500$}
        %\label{K64_M128_vc}
        % \addplot[BurntOrange, mark=diamond]
        % table[x=R,y=MMSE_cen_inff,col sep=comma] {data/revision1/data_diffR.csv};
        %\addlegendentry{$b=200$}
% \end{axis}
% \end{scope}

% \begin{scope}[xshift=0,yshift=-6cm]
% \begin{axis}[
% domain=0:4,
% grid=major,
% xlabel= {Total fronthaul capacity (bit/uplink sample)},
% xmin=100,
% xmax=1000,
% xtick style={color=clr3},
% xticklabel style={yshift=-2.5pt},
% %xticklabels={100,200,300,400,500,600,700,800,900,1000},
% ymin=0,
% ymax=40,
% mark size=4.0 pt,
% legend columns=3,
% legend style={nodes={scale=0.9, transform shape},at={(0.5,-0.25)},anchor=north},
% line width=1pt,
% % yticklabels=\empty,
% ylabel= {Average sum SE (bit/sec/Hz)},
% %ylabel= Average sum spectral efficiency,
% ]

         \addplot [blue]
        table[x=R,y=eq_p2_eq,col sep=comma] {data/revision1/data_diffR.csv};
        \addlegendentry{EF-EIU}

        \addplot [BurntOrange]
        table[x=R,y=MSE_p2_eq,col sep=comma] {data/revision1/data_diffR.csv};
         \addlegendentry{EF-SCNM}

        \addplot [red]
        table[x=R,y=WMSE_p2_eq,col sep=comma] {data/revision1/data_diffR.csv};
         \addlegendentry{EF-WSINM}

         \addplot [blue, dashed]
        table[x=R,y=eq_p2_lin,col sep=comma] {data/revision1/data_diffR.csv};
        \addlegendentry{LF-EIU}

        \addplot [BurntOrange, dashed]
        table[x=R,y=MSE_p2_lin,col sep=comma] {data/revision1/data_diffR.csv};
         \addlegendentry{LF-SCNM}

        \addplot [red, dashed]
        table[x=R,y=WMSE_p2_lin,col sep=comma] {data/revision1/data_diffR.csv};
         \addlegendentry{LF-WSINM}
\draw[ black] (600,10) ellipse [x radius=20, y radius=4];
\draw[inner sep=50mm, black] (600,25) ellipse [x radius=20, y radius=6];
    \draw[->] (600,10) -- (750,10) node[right] {Single-path};    
    \draw[->] (600,25) -- (450,25) node[left] {Two-path};
\end{axis}
\end{scope}

\end{tikzpicture}

%\end{document}
    \caption{Impact of limited capacity fronthaul link: sum SE vs total fronthaul capacity $R_T$. $L=12$, $M=120$ and $K=20$. 
    % Left: Single-Path signal estimation, Right: Two-Path signal estimation.
    }
    \label{fig4_diffR}
\end{figure}
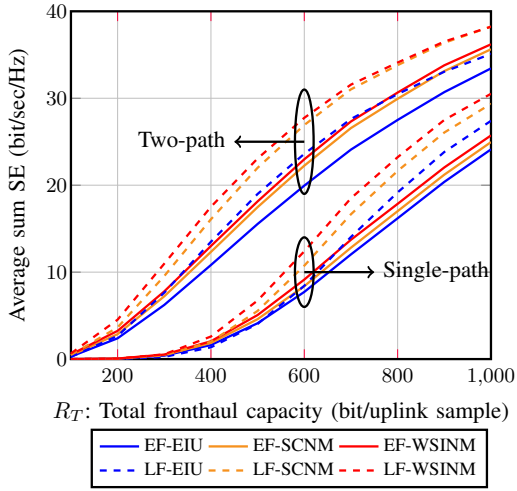

Having two paths for sequential signal processing and fusion in the CPU can mitigate the adverse effect of fronthaul compression on the sum users' SE, even with the most rudimentary fronthaul capacity allocation among APs (i.e., EF) and among users (i.e., EIU), as shown in Fig. \ref{fig4_diffR}. 

\section{Conclusion}

This paper shows that given a total fronthaul capacity in the cell-free massive MIMO (CFmMIMO) networks with sequential fronthaul, e.g., a total number of subcarriers allocated to perform wireless fronthaul signaling, the APs toward the end of the sequence should get a larger share of this total fronthaul capacity than the APs at the beginning of the sequence. The heterogeneous fronthaul capacity allocation enhances user spectral efficiency (SE), especially when the number of users is high. Ultimately, breaking the Single-Path signal estimation into Two-Path signal estimation can alleviate the detrimental effect of fronthaul compression, even when the equal fronthaul allocation among APs (EF) is considered. It is worth noting that the proposed strategies can be extended to other fronthaul topologies, as many fronthaul topologies can be reduced to a virtual sequential topology with virtual nodes (APs or clusters of APs) connected sequentially to each
other. Future work can look into optimizing the number of APs per path in Two-Path signal propagation, and optimizing the fronthaul capacity allocation among APs instead of using heuristic approaches.
%The motivation for using logarithmic allocation is that SINR enhancement is a logarithmic function of the number of APs (or number of antennas per AP). However, this logarithmic function saturates more slowly as the number of users increases.s a result, for a larger number of users, we observe that linear allocation outperforms the proposed logarithmic allocation.  

% \section*{Acknowledgments}
% This work is supported by the European Union’s Horizon Europe research and
% innovation programme under Grant Agreement No 101096954 (6G-BRICKS) and by the Research Foundation – Flanders (FWO) under
% project number G0C0623N. The resources and services used in this work were
% provided by the VSC (Flemish Supercomputer Center), funded by the Research
% Foundation - Flanders (FWO) and the Flemish Government.
{%\appendices
\appendix[Proof that the objective function in \ref{min_intf_prob_VC} is bounded below.]
The objective function in (\ref{min_intf_prob_VC}) is convex with respect to $\{w_{lk}, \forall k\}$. What follows proves that the objective function is a summation of $K$ convex and bounded from below functions. For the sum of such functions, convex and bounded properties are preserved.
Hence, this chapter proceeds to prove that each of the summands is convex and bounded from below. For the sake of simplicity, the subscripts are removed. Function $f$ defined as
\begin{equation}
    f=wX-\log_2(w)
\end{equation}
is convex with respect to $w$. This is proved by showing that the second derivative of $f$ with respect to $w$ is positive. The first and second derivatives of the function are formulated as follows:
\begin{equation}
    \begin{aligned}
        \frac{df}{dw}&=X-\frac{1}{\ln2\times w}
    \end{aligned}
\end{equation}
and
\begin{equation}
    \begin{aligned}
        \frac{d^2f}{{dw^2}}&=\frac{1}{\ln2\times w^2}
    \end{aligned}
\end{equation}
The second derivative is always positive, and hence, the convexity of the function is proved. The minimum of the function occurs at the point where the derivative is zero:
\begin{equation}
    \begin{aligned}
        X-\frac{1}{\ln2\times w^*}=0\rightarrow w^*=\frac{1}{\ln2\times X}
    \end{aligned}
\end{equation}
% \robbert{The global minimum of the function can then be calculated in closed form by substituting ... .}
The minimum value of the function is then calculated in closed form by substituting 
$w$ in $f$ with $w^*$, as follows:
\begin{equation}
    \begin{aligned}
        f^*&=w^*X-\log_2(w^*)\\&=\frac{1}{\ln2}-\log_2(\frac{1}{\ln2 \times X})\\&=\frac{1}{\ln2}-(\log_2(\frac{1}{\ln2})+\log_2(\frac{1}{X}))\\&=\frac{1}{\ln2}+\log_2({\ln2})+\log_2({X})
    \end{aligned}
\end{equation}
For all $X>0$, the value $f^*>-\infty$. It is almost certain that $X>0$, i.e., the interference plus noise term is non-zero for any user, especially for a multi user scenario ($K>$). $X$ has a non-zero and positive value in a given simulation setup, and this value defines the bound of the function from below. 
% \robbert{Since it is reasonable to assume that the interference is non-zero for any system with $K > 1$, the global minimum of the convex function $f$ is bounded from below, and hence the entire function is bounded from below. }
Fig. (\ref{ch5_app:figC1}) shows Function $f$ in 3D for illustration purposes.
\begin{figure}
    \centering
    \includegraphics{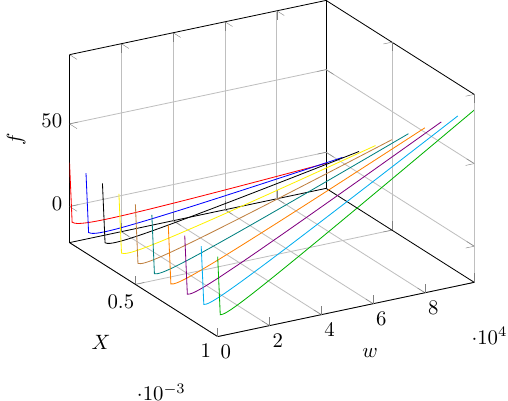}
    \caption{3D plot of $f$ as a function of $w$ for different values of $X$.}
    \label{ch5_app:figC1}
\end{figure}
% The objective function in problem (\ref{min_intf_prob_VC}) is the summation of functions similar to $f$ with different $X$ values, some of which are shown in Fig. \ref{ch5_app:figC1}.
To illustrate our point, we have generated several realizations of $f$ for different values of $X$ in Fig.\ref{fig4_diffR}. It is clear from the figure that $f$ is bounded from below, as is the objective function in (\ref{min_intf_prob_VC}).
%\bibliographystyle{ieeetr}
%\bibliography{refs}
%%%%%%%%%%%%%%%%%%%%%%%%%%%%%%%%%%%%%%%%%%%%%%%%%%
% Keep the following \cleardoublepage at the end of this file, 
% otherwise \includeonly includes empty pages.
\bibliographystyle{IEEEtran}
\bibliography{bibliography}

@inproceedings{VRAN_eucnc_2024,
	title        = {{Joint Sequential Fronthaul Quantization and Hardware Complexity Reduction in Uplink Cell-Free Massive MIMO Networks}},
	author       = {Ranjbar, Vida and Beerten, Robbert and Moonen, Marc and Pollin, Sofie},
	year         = {2024},
	booktitle    = {2024 Joint European Conference on Networks and Communications \& 6G Summit (EuCNC/6G Summit)},
	volume       = {},
	number       = {},
	pages        = {1--6},
	doi          = {10.1109/EuCNC/6GSummit60053.2024.10597050}
}

@article{shaik2021,
	title        = {{MMSE-Optimal Sequential Processing for Cell-Free Massive MIMO With Radio Stripes}},
	author       = {Shaik, Zakir Hussain and Bj{\"o}rnson, Emil and Larsson, Erik G.},
	year         = {2021},
	journal      = {IEEE Transactions on Communications},
	volume       = {69},
	number       = {11},
	pages        = {7775--7789},
	doi          = {10.1109/TCOMM.2021.3100619}
}

@inproceedings{shaik2020,
	title        = {{Cell-Free Massive MIMO with Radio Stripes and Sequential Uplink Processing}},
	author       = {Shaik, Zakir Hussain and Bj{\"o}rnson, Emil and Larsson, Erik G.},
	year         = {2020},
	booktitle    = {IEEE International Conference on Communications Workshops (ICC Workshops)},
	volume       = {},
	number       = {},
	pages        = {1--6},
	doi          = {10.1109/ICCWorkshops49005.2020.9145164}
}

@book{elgamal_kim_2011,
	title        = {{Network Information Theory}},
	author       = {El Gamal, Abbas and Kim, Young-Han},
	year         = {2011},
	publisher    = {Cambridge University Press},
	doi          = {10.1017/CBO9781139030687},
	place        = {Cambridge}
}

@article{3gpp_PL,
	title        = {{Further advancements for E-UTRA physical layer aspects (Re- lease 9)}},
	author       = {3GPP},
	year         = {Mar. 2017},
	journal      = {3GPP TS 36.814},
	volume       = {},
	number       = {},
	pages        = {},
	doi          = {}
}

@inproceedings{ke_Dmimo_kalman,
	title        = {{Uplink D-MIMO Processing Using Kalman Filter Combining}},
	author       = {Helmersson, Ke Wang and Frenger, Pål and Helmersson, Anders},
	year         = {2022},
	booktitle    = {IEEE Global Communications Conference},
	volume       = {},
	number       = {},
	pages        = {1703--1708},
	doi          = {10.1109/GLOBECOM48099.2022.10001527}
}

@article{massivemimobook,
	title        = {{Massive {MIMO} Networks: {Spectral}, Energy, and Hardware Efficiency}},
	author       = {Emil Bj\"{o}rnson and Jakob Hoydis and Luca Sanguinetti},
	year         = {2017},
	journal      = {Foundations and Trends{\textregistered} in Signal Processing},
	volume       = {11},
	number       = {3-4},
	pages        = {154--655},
	doi          = {10.1561/2000000093},
	issn         = {1932-8346},
	url          = {http://dx.doi.org/10.1561/2000000093}
}

@book{mcook,
	title        = {{The Matrix Cookbook}},
	author       = {K. B. Petersen and M. S. Pedersen},
	year         = {2012},
	month        = {},
	pages        = {},
	isbn         = {},
        note  = {\url {https://www.math.uwaterloo.ca/~hwolkowi/matrixcookbook.pdf}}
}

@article{Ioaniss_TWC_2024,
	title        = {{Uplink Performance Optimization of Limited-Capacity Radio Stripes}},
	author       = {Chiotis, Ioannis and Moustakas, Aris L.},
	year         = {2024},
	journal      = {IEEE Transactions on Wireless Communications},
	volume       = {23},
	number       = {9},
	pages        = {12382--12395},
	doi          = {10.1109/TWC.2024.3392178}
}

@inproceedings{Ioanis_meditcom_2022,
	title        = {{On the Uplink Performance of Finite-Capacity Radio Stripes}},
	author       = {Chiotis, Ioannis and Moustakas, Aris L.},
	year         = {2022},
	booktitle    = {IEEE International Mediterranean Conference on Communications and Networking (MeditCom)},
	volume       = {},
	number       = {},
	pages        = {118--123},
	doi          = {10.1109/MeditCom55741.2022.9928628}
}

@inproceedings{Ioanis_ISNCC_2023,
	title        = {{Optimal MMSE Processing for Limited-Capacity Radio Stripes}},
	author       = {Chiotis, Ioannis and Moustakas, Aris L.},
	year         = {2023},
	booktitle    = {International Symposium on Networks, Computers and Communications (ISNCC)},
	volume       = {},
	number       = {},
	pages        = {1--6},
	doi          = {10.1109/ISNCC58260.2023.10323670}
}

@inproceedings{Ioanis_spawc_2024,
	title        = {{A Sequential UE Selection Scheme for User-Centric Communications Employing Radio Stripes}},
	author       = {Chiotis, Ioannis and Moustakas, Aris L.},
	year         = {2024},
	booktitle    = {IEEE 25th International Workshop on Signal Processing Advances in Wireless Communications (SPAWC)},
	volume       = {},
	number       = {},
	pages        = {161--165},
	doi          = {10.1109/SPAWC60668.2024.10694488}
}

@inproceedings{Jo_joint_prc_comp,
	title        = {{Joint Precoding and Fronthaul Compression for Cell-Free MIMO Downlink With Radio Stripes}},
	author       = {Jo, Sangwon and Lee, Hoon and Park, Seok-Hwan},
	year         = {2023},
	booktitle    = {IEEE Global Communications Conference},
	volume       = {},
	number       = {},
	pages        = {4945--4951},
	doi          = {10.1109/GLOBECOM54140.2023.10437610}
}

@inproceedings{Ioannis_2024_MI,
	title        = {{Mutual Information Optimization for Cell-Free Massive MIMO with Sequential Limited-Rate Fronthaul}},
	author       = {Chiotis, Ioannis and Moustakas, Aris L.},
	year         = {2024},
	booktitle    = {IEEE Middle East Conference on Communications and Networking (MECOM)},
	volume       = {},
	number       = {},
	pages        = {145--150},
	doi          = {10.1109/MECOM61498.2024.10881772}
}

@ARTICLE{wmse_2011,
  author={Shi, Qingjiang and Razaviyayn, Meisam and Luo, Zhi-Quan and He, Chen},
  journal={IEEE Transactions on Signal Processing}, 
  title={{An Iteratively Weighted MMSE Approach to Distributed Sum-Utility Maximization for a MIMO Interfering Broadcast Channel}}, 
  year={2011},
  volume={59},
  number={9},
  pages={4331-4340},
  doi={10.1109/TSP.2011.2147784}}

@ARTICLE{beamsync,
  author={Kunnath Ganesan, Unnikrishnan and Sarvendranath, Rimalapudi and Larsson, Erik G.},
  journal={IEEE Transactions on Wireless Communications}, 
  title={BeamSync: Over-the-Air Synchronization for Distributed Massive MIMO Systems}, 
  year={2024},
  volume={23},
  number={7},
  pages={6824-6837},
  doi={10.1109/TWC.2023.3335089}}

@article{zhang2020downlink,
  title        = {On the Downlink Capacity of Cell-Free Massive {MIMO} with Constrained Fronthaul Capacity},
  author       = {Zhang, Peng and Willems, Frans M. J.},
  journal      = {Entropy},
  year         = {2020},
  volume       = {22},
  number       = {4},
  pages        = {418},
  doi          = {10.3390/e22040418},
}

\vfill

\end{document}